\documentclass{article}
\usepackage{spconf,amsmath,graphicx}
\usepackage{graphicx}
\usepackage{epsfig}
\usepackage{graphicx}
\usepackage{amsmath}
\usepackage{amssymb}
\usepackage{multirow}
\usepackage{tabu}
\usepackage{booktabs}
\usepackage{bbm}
\usepackage{diagbox}
\usepackage[english]{babel}
\usepackage{comment}

\usepackage{xcolor}
\usepackage{subfigure}
\usepackage{wrapfig}
\usepackage{cite}
\usepackage{diagbox}
\usepackage{array}
\usepackage{amstext}
\usepackage{makecell}
\usepackage{caption}
\usepackage{tablefootnote}
\usepackage{babel}

\usepackage{amsmath}
\usepackage{chngcntr}
\usepackage{placeins}
\usepackage{booktabs}       
\usepackage{amsfonts}       
\usepackage{nicefrac}  

\title{Relational Learning between Multiple Pulmonary Nodules via Deep Set Attention Transformers}
\name{
Jiancheng Yang\textsuperscript{1,2,3} \quad
Haoran Deng\textsuperscript{1} \quad
Xiaoyang Huang\textsuperscript{1} \quad
Bingbing Ni\textsuperscript{1,2,4}\sthanks{Corresponding author: Bingbing Ni.} \quad
Yi Xu \textsuperscript{1}}

\address{
\textsuperscript{1}Shanghai Jiao Tong University, Shanghai, China\\
\textsuperscript{2}MoE Key Lab of Artificial Intelligence, Artificial Intelligence Institute, Shanghai Jiao Tong University\\
\textsuperscript{3}Diannei Technology, Shanghai, China \quad \textsuperscript{4}Huawei Hisilicon\\
{\tt\small \{jekyll4168,dhrdeng,13802896410,nibingbing,xuyi\}@sjtu.edu.cn}}

\begin{document}

\maketitle

\begin{abstract}

Diagnosis and treatment of multiple pulmonary nodules are clinically important but challenging. Prior studies on nodule characterization use solitary-nodule approaches on multiple nodular patients, which ignores the relations between nodules. In this study, we propose a multiple instance learning (MIL) approach and empirically prove the benefit to learn the relations between multiple nodules. By treating the multiple nodules from a same patient as a whole, critical relational information between solitary-nodule voxels is extracted. To our knowledge, it is the first study to learn the relations between multiple pulmonary nodules. Inspired by recent advances in natural language processing (NLP) domain, we introduce a self-attention transformer equipped with 3D CNN, named {NoduleSAT}, to replace typical pooling-based aggregation in multiple instance learning. Extensive experiments on lung nodule false positive reduction on LUNA16 database, and malignancy classification on LIDC-IDRI database, validate the effectiveness of the proposed method.
\end{abstract}
\begin{keywords}
Relational Learning, Attention, Multiple Pulmonary Nodules, Computer-Aided Diagnosis (CADx).
\end{keywords}
\section{Introduction}

Lung cancer is the leading cause of cancer-related mortalities. Early diagnosis of lung cancer is one of the effective ways to reduce the related death. With prevailing application of low-dose computed tomography (LDCT), increasing number of early-stage pulmonary nodules have become a challenge in clinical practice. In particular, a set of subjects with {\em multiple pulmonary nodules} raises research attention \cite{sobue2002screening}. Except for certain ``easy-to-diagnose" diseases, e.g., evident pulmonary metastases and pulmonary tuberculosis, incidental multiple pulmonary nodules are considered as a dilemma in clinical context. Diagnosis of multiple pulmonary nodules is more complicated than solitary ones; apart from analyzing the biological behaviors (e.g., benign, indolent, invasive), radiologists are required to analyze diverse circumstances. 
 
Recent data-driven approach, e.g., radiomics and deep learning, has been dominating Computer-Aided Diagnosis (CADx) research. Few prior study on lung nodule detection and characterization works on learning the relations between multiple nodules; in other words, prior studies use solitary-nodule approaches on multiple nodular patients. In clinical practice, radiologists use information at nodule level and patient level to diagnose nodules from the same subject; from an algorithmic perspective, solitary-nodule approaches classify nodules without considering relation / context information. We argue that {\bf relation does matter}, thereby a multiple instance learning (MIL) approach is proposed to address this issue. To our knowledge, it is the first study to learn the relations between multiple pulmonary nodules. Inspired by the recent advances in NLP domain \cite{vaswani2017attention}, we introduce {\em Set Attention Transformers (SATs)} based on self-attention to learn the relations between nodules, instead of typical pooling-based aggregation in multiple instance learning. To model lung nodules from CT scans, a 3D DenseNet \cite{huang2017densely, zhao20183d,yang2019probabilistic,zhao2019toward} is used as backbone to learn representations of voxels at solitary-nodule level. We then use the SATs to learn the relations between multiple nodules from the same patient. The whole network, named \textit{NoduleSAT}, could be trained end-to-end. In lung nodule false positive reduction and malignancy classification tasks, the proposed multiple-nodule approach consistently outperforms the solitary-nodule baselines.

The key contributions of this paper are threefold: 1) We empirically prove the benefit to learn the relations between nodules; 2) We develop Set Attention Transformers (SATs) to perform the relational learning; 3) An end-to-end trainable NoduleSAT is shown effective in modeling multiple nodules. We empirically prove the benefit of relation learning between multiple pulmonary nodules, which could also inspire clinical and biological research on this important topic.

\section{Methodology}

\subsection{Set Attention Transformers (SATs)}

Inspired by the self-attention transformers \cite{vaswani2017attention}, we develop a Set Attention Transformer (SATs), a general module designed for processing the is permutation-invariant and size-varying \textit{set}. In this study, the set is the multiple nodules from a same patient, and the SAT is designed to learn the relations between the multiple pulmonary nodules. For an input set $X \in \mathbb{R}^{N\times c}$ ($N$ denotes the size of the set, and $c$ denotes the channels of representation), a scaled dot-product attention is formulated by sharing the $K$, $Q$ and $V$ features \cite{vaswani2017attention},

\begin{equation} \label{eq:non-linear-attn}
\mathit{Attn}(X) = \mathit{softmax}(XX^T/\sqrt{c})\cdot \sigma(X),
\end{equation}
where $\sigma$ is a non-linearity activation function.

However, the \textit{Multi-Head Attention} (MHA) \cite{vaswani2017attention} is too much ponderous in our application, we introduce a more parameter-efficient {\em Group Shuffle Attention (GSA)} \cite{Yang_2019_CVPR} for SATs. Suppose $g$ to be the number of groups ($g=8$ in this study), $c_g = c/g$, $s.t. \; c\mod g =0$, we divide $X$ by channels into $g$ groups: $\{X^{(i)} \in \mathbb{R}^{N \times c_g}\}$, and apply group linear transform by the weight $W_g \in \mathbb{R}^{c_g \times c_g}$, before the scaled dot-product attention (Eq. \ref{eq:non-linear-attn}). In-group scaled dot-production attention is applied in each group. However, grouping the inputs in all layers results in the no communication between the elements in the sets. For this reason, a parameter-free operator, channel shuffle $\psi$ \cite{Yang_2019_CVPR}, is introduced to encourage channel fusion. The overall formalization of \textit{GSA} is as follows, 

\begin{scriptsize}
\begin{equation}
\mathit{GSA}(X)
= \mathcal{BN}(\psi(\mathit{concat}\{\mathit{Attn}(X_i)|X_i=X^{(i)}W_i\}_{i=1,..,g})+X),
\end{equation}
\end{scriptsize}
where Batch Normalization $\mathcal{BN}$ \cite{ioffe2015batch} is introduced to ease the optimization. The parameter size of \textit{GSA} is up to $4g\times$ smaller than \textit{MHA}, which makes the SATs very light-weight and easy to learn. An SAT is simply an $L$-layer stack of GSA operator. 

\subsection{3D DenseNet Backbone} \label{sec:DenseNet-config}
To end-to-end learn the representation of lesion voxels from CT scans, we apply a 3D DenseNet \cite{huang2017densely, zhao20183d,yang2019probabilistic,zhao2019toward} as backbone, with a compression rate $\theta=2$ and a bottleneck $B=4$. Leaky ReLU ($\alpha=0.1$) together with Batch Normalization \cite{ioffe2015batch} are used as activation functions. We instantiate various DenseNets for different experiments (Sec \ref{sec:experiments}).

Before inputting into the 3D DenseNet, the nodule voxels from CT scans are pre-processed with the following procedure: 1) clip the Hounsfield units into $[-1024,400]$, 2) linearly transformed into $[-1,1]$, and 3) resize the volumetric data into a spacing of $1mm\times 1mm \times 1mm$ with trilinear interpolation. We also apply online data augmentation including $90^{\circ},180^{\circ},270^{\circ}$ rotation in random axis, left-right flipping and shifting the voxel centers in $[-1,1]$.

\begin{figure}
    \centering
	\includegraphics[width=0.45\textwidth]{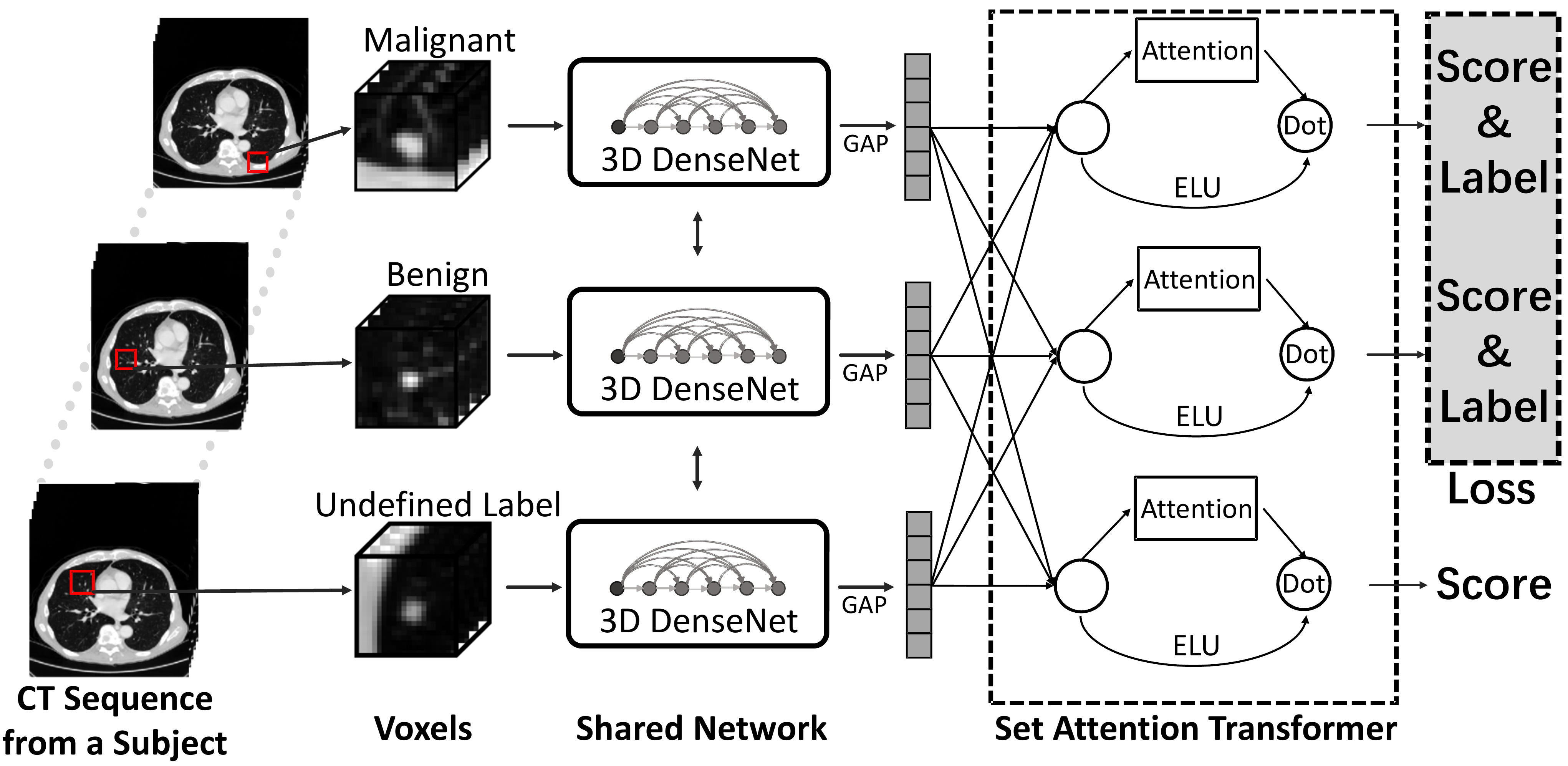}
	\caption{{\bf Model architecture of NoduleSATs.} We demonstrate the multiple pulmonary nodule malignancy classification task on the figure. Volumes of multiple pulmonary nodules from the same subject (patient) are input for a shared 3D DenseNet, to produce nodule-level representations via global average pooling (GAP). A Set Attention Transformer (SAT) is used for learning the relations between the multiple nodules, and it outputs a refined representation for each nodule. For loss back-propagation, we only accumulate the classification loss for the nodules with ``malignant" and ``benign" labels, and ignore the nodules with ambiguous labels. } 

	\label{fig:nodule-sat}
\end{figure}

\subsection{End-to-end NoduleSAT}

Combining the SATs and 3D DenseNets, the proposed network for learning the relations between multiple pulmonary nodules is named NoduleSAT (\ref{fig:nodule-sat}). We treat the multiple pulmonary nodules from a same subject (patient) as a set, and use a shared-weight 3D DenseNet \cite{huang2017densely, zhao20183d,yang2019probabilistic,zhao2019toward} to extract nodule-level representation for each nodule. These representations are fed into an $L$-layer SAT with hidden size $H$ to learn the relations. Note that the whole NoduleSAT network could be trained end-to-end by standard back-propagation. Experiments on nodule false positive reduction (Sec. \ref{sec:lung-nodule-fpr}) and nodule malignancy classification (Sec. \ref{sec:nodule-malignancy}) are conducted to validate the usefulness of the proposed method. For both tasks, the NoduleSAT is trained to classify whether the multiple candidates / nodules from a same patient are nodules / malignant or not. Specifically for the nodule malignancy classification task, to take advantage of the nodules with ambiguous / undefined labels, we design a {\em masked BCE loss} to train the NoduleSATs. For solitary-nodule approaches, these data samples are non-trivial to use. In our NoduleSAT framework, these nodules are ignored in the loss back-propagation, but presented as input to provide important context information to other nodules. The masked loss strategy extend the effective sample size, and is shown to be effective in our experiments (Sec \ref{sec:nodule-malignancy}).

It is notable that the proposed NoduleSAT network is fundamentally different from a prior study using multiple instance learning on lung nodules \cite{liao2019evaluate}, where a max-pooling aggregate multiple-nodule information and output a single patient-level representation. Thereby, no relational information between the multiple nodules could be captured. As a comparison, relational information could be learned via layer-by-layer self-attention in the proposed NoduleSAT.

\section{Experiments} \label{sec:experiments}

\subsection{Lung Nodule False Positive Reduction} \label{sec:lung-nodule-fpr}

\subsubsection{Dataset}

False positive reduction (FPR) in lung nodule detection has always been a demanding task in computer-aided detection research. Due to the objectively existing correlation and largely varied size of nodules from a subject, our SAT is more than suitable to address this problem.

We use two datasets for this experiment. One is LUNA16, a widely used dataset for lung nodule detection and false positive reduction. The LUNA16 dataset consists of CT scans of 888 subjects with 1186 nodules. The evaluation result is provided as 10-fold cross validation using the official split. The LUNA16 FPR dataset uses the candidate list provided by the competition host, totally $754,975$ candidates. We filter out the candidates with a predicted score (by our 3D DenseNets baseline) lower than $0.1$, resulting in $32,405$ candidates for further refinement by our SATs. The second dataset is Tianchi Lung Nodule Detection dataset\footnote{https://tianchi.aliyun.com/competition/entrance/231601/introduction}, a dataset of similar data protocol as LUNA16, while it consists of $800$ subjects with $1,224$ nodules in total. We use the official split, with $600$ subjects ($975$
nodules) for training and $200$ subjects ($269$ nodules) for validation.
The performance are reported on the validation set (named Tianchi VAL). We use our nodule detection model (based on a 3D UNet) trained on training set, to extract the candidates on the CT scans, resulting in $5,531$ candidates on the training set, $1,515$ on the validation set.

\begin{table}[!htb]
\renewcommand\arraystretch{0.5}
 		\centering
 		\caption{Lung nodule false positive reduction performance.}
		\begin{tabular}{lcc}
			\toprule
			Dataset & Method &  Average FROC (CPM)  \\
			\midrule
			\multirow{4}{*}{LUNA16 \cite{setio2017validation}} &
			2D-CNN \cite{xie2019automated} &  0.790 \\
			& 3D-CNN \cite{dou2017multilevel}  & 0.908\footnotemark  \\
			& 3D DenseNet & 0.884 \\
			& NoduleSAT  & {\bfseries 0.916} \\
			\midrule
			\multirow{2}{*}{Tianchi VAL} & 3D DenseNet &  0.677    \\
			& NoduleSAT & {\bfseries 0.716} \\
			\bottomrule
		\end{tabular}
	
	\label{nodule-fpr}
\end{table}
 \footnotetext{We refer to their updated result (https://luna16.grand-challenge.org/), the publication paper result is 0.827.}

\subsubsection{Experiment Setting}
A 3D CNN based on DenseNets \cite{huang2017densely} is used as a strong baseline. The input size is $48\times48\times48$. At each resolution level, dense blocks are repeated $[4,4,4,4]$ times before each down-sampling, with a growth rate of $16$. We use an Adam optimizer for training the CNN with a batch size of $32$, with an initial learning rate $1\times 10^{-3}$. We exponentially decay the learning rate with a ratio of $3\times 10^{-2}$ after every epoch. The candidates from the same subject, represented by the features after the global average pooling of CNN, are fed together into an SAT. Since there are too many candidates on certain subjects, we fix the pre-trained 3D DenseNet to train the SAT.

The SAT in the false positive reduction experiments uses $H=256$ and $L=3$. We use an Adam optimizer with a batch size of $64$. The initial learning rate is $5\times 10^{-3}$. We multiply the learning rate by $0.2$ at $epoch=100, 130$ and $170$. 200 epochs are enough for a good convergence. The training loss is a cross entropy loss averaged by the number of candidates.

\subsubsection{Result}

We use CPM score for evaluation, the most commonly used metric for lung nodule detection and false positive reduction.
CPM score is the average recall rate at an average number of false positives at $0.125, 0.25, 0.5, 1.0, 2.0, 4.0$ and $8.0$ on the FROC curve, the higher is better.

Results are shown in Table \ref{nodule-fpr}, our SAT-based method improves the baseline by a large margin. On Tianchi VAL, our model improves more significantly. Note that a 10-run voting ensemble usually provides only $1\%$ performance boost on this dataset.
We declare that these improvements come from {\bfseries learning the relations} between the candidates. 

\subsection{Multiple Nodule Malignancy Classification}\label{sec:nodule-malignancy}
\subsubsection{Dataset}
We then use the NoduleSAT to provide a systemic view on malignancy of multiple pulmonary nodules. LIDC-IDRI \cite{armato2011lung}, one of the largest public available lung cancer screening databases, is used for validating our method.
As depicted in Fig \ref{fig:subfig} (a), patients in LIDC-IDRI dataset have 1 - 23 nodules, of which $74.0\%$ are multiple nodular patients. Therefore, the proposed NoduleSAT network is well suitable for this task. 
The data inclusion criteria are: 1) the nodule should be annotated by at least 3 radiologists, and 2) the CT thickness$\leq$ 3mm. 2,175 qualified nodules are selected in total; we then compute the average malignancy score ($s_{avg}$) of each nodule, resulting in 527 malignant ($s_{avg}>3$), 656 benign ($s_{avg}<3$) and 992 undefined-label (or ambiguous, $s_{avg}=3$) nodules.

\subsubsection{Experiment Setting}
We first pre-train a 3D DenseNet with a $32\times32\times32$ input. The dense blocks are repeated $[3,8,4]$ times before each down-sampling, with a growth rate $k=32$. The training samples are only the 1,183 benign or malignant nodules. The 3D DenseNet is trained with a standard cross entropy loss. We train the 3D DenseNet using an Adam optimizer with an initial learning rate of $0.001$ for $200$ epochs, and halve the learning rate every 30 epochs. 

We then construct a NoduleSAT, using the pre-trained 3D DenseNet and a $3$-layer SAT ($H=256$). We train the NoduleSAT end-to-end with a batch of 16 patients. Note the batch size for 3D DenseNet is variable. We fix the Batch Normalization layer to stabilize the training. All qualified 2,175 nodules are used for training the NoduleSAT, with a masked loss to backward on benign and malignant nodules only. We train the whole NoduleSAT with an Adam optimizer, whose learning rate is initially $0.001$, halved every 15 epochs for 150 epochs totally. Another NoduleSAT with 1,183 benign or malignant nodules are also trained for fair comparison.

\subsubsection{Result}

\begin{figure}[!htb]

  \centering
  \subfigure[Nodule Count Distribution]{
    \label{nodule_count} 
    \includegraphics[width=0.45\linewidth]{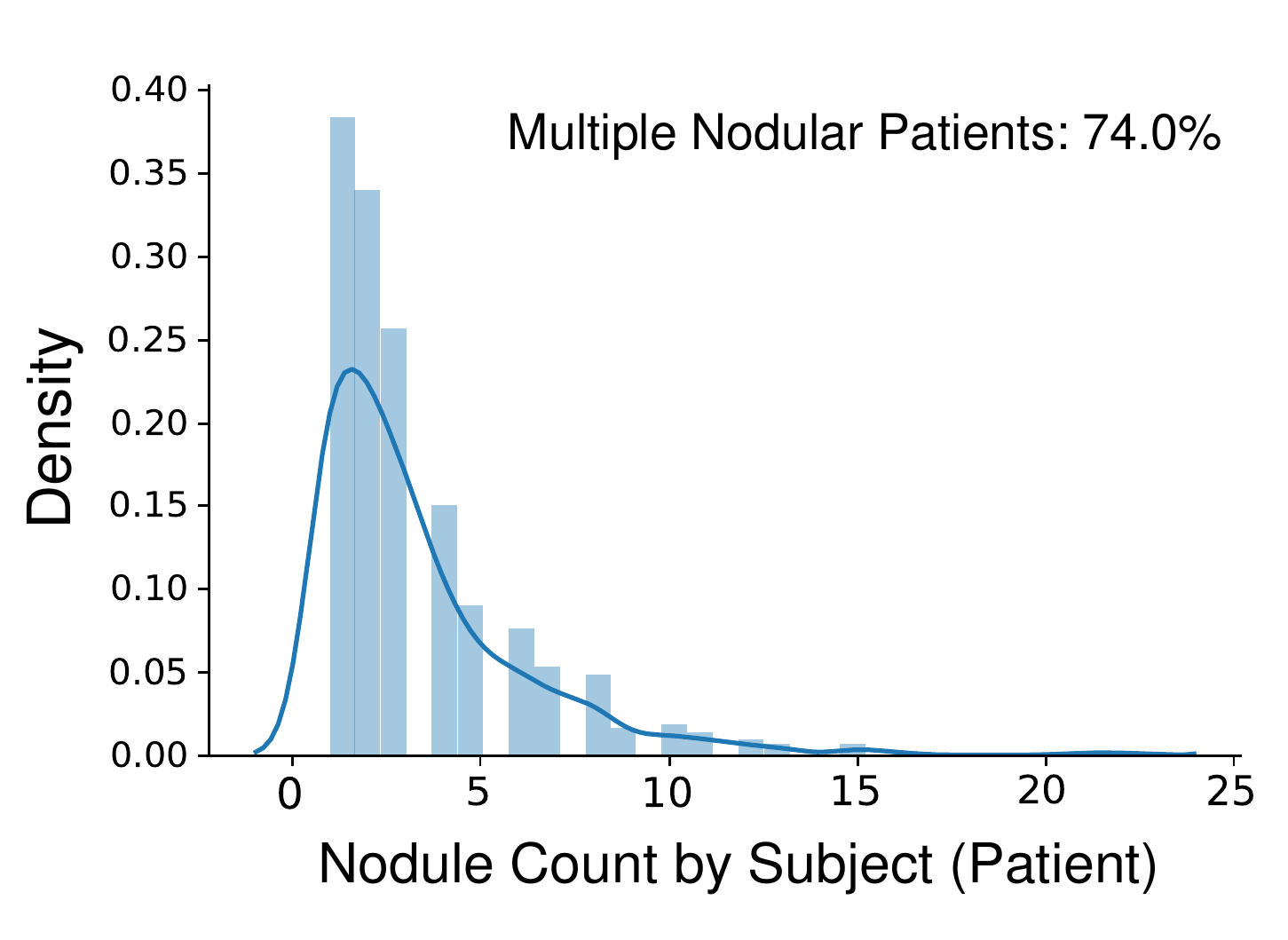}}
  \subfigure[Model Performance]{
    \label{model_performance_on_LIDC-IDRI} 
    \includegraphics[width=0.45\linewidth]{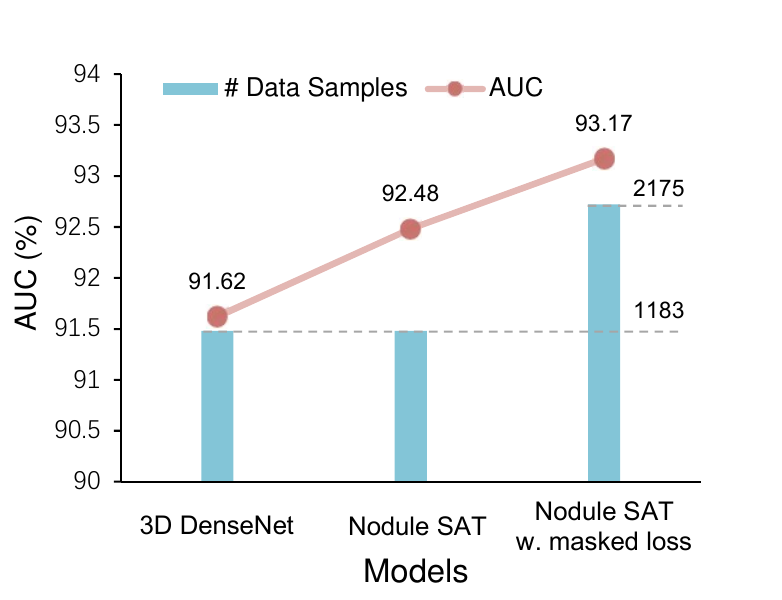}}
    \vspace{-10px}
    
    \caption{(a) The distribution and kernel density estimation of nodule count by subject (patient). (b) Model performance on LIDC-IDRI nodule malignancy classification with the baseline 3D DenseNet and NoduleSAT. The blue bar denotes the effective data samples with the corresponding method, and the red curve denotes the AUC with various settings.}
    \vspace{-10px}
  \label{fig:subfig} 
\end{figure}

We report the AUC (AUROC) with a 5-fold cross validation for evaluating our method, where each fold is split by the patients, while keeping roughly the same number of nodules in each fold. As depicted in Figure \ref{fig:subfig} (b), NoduleSAT and the masked loss are both effective to boost malignancy classification performance from the baseline 3D DenseNet. We figure out two important findings: 1) NoduleSAT with only the benign or malignant nodules outperforms 3D DenseNet with the same dataset, and we attribute the improvement to {\bf learning the relations}. 2) NoduleSAT with the undefined-label nodules, using masked loss, boosts the performance further. We argue that this improvement comes from {\bf learning the context}. Our method elegantly uses the undefined-label samples, which is non-trivial for solitary-nodule approaches.

\section{Conclusion and Further Work}

In this study, we propose a Set Attention Transformer (SAT), to explicitly learn relational information between multiple pulmonary nodules from a same subject. Intergated with a 3D DenseNet, the proposed end-to-end trainable NoduleSAT encourages the model to learn top-down inter-nodule relations from bottom-up nodule-level representations.

We are working on clinical problems on multiple pulmonary nodules. Hopefully, our data-driven methodology could benefit understanding etiologic and biologic processes and metastasis diagnosis of multiple pulmonary nodules.

\small{
\noindent\textbf{Acknowledgment.}
The authors would like to thank Dr. Wei Zhao (Huadong Hospital) and Guangyu Tao (Shanghai Chest Hospital) for insightful discussion. This work was supported by National Science Foundation of China (61976137, U1611461). This work was also supported by Interdisciplinary Program of Shanghai Jiao Tong University (YG2017QN661), SJTU-BIGO Joint Research Fund and SJTU-UCLA Joint Research center.
}

\bibliographystyle{IEEEbib}
\bibliography{refs}

\end{document}